\documentclass[fleqn,usenatbib]{rasti}
    
\usepackage[T1]{fontenc}
\usepackage{ae,aecompl}

\usepackage{graphicx}	
\usepackage{amsmath}	
\usepackage{amssymb}	
\usepackage{multirow}

\usepackage{caption}
\usepackage{subcaption}
\usepackage{newtxtext,newtxmath}

\DeclareRobustCommand{\VAN}[3]{#2}
\let\VANthebibliography\thebibliography
\def\thebibliography{\DeclareRobustCommand{\VAN}[3]{##3}\VANthebibliography}

\newcommand{\ampere}{{\sc ampere}} 

\title[\sc ampere]{\ampere: A tool to fit heterogeneous observations consistently}

\author[P. Scicluna et al.]{P. Scicluna$^{1, 2, 3}$\thanks{E-mail: p.scicluna@herts.ac.uk}, S. Zeegers$^{4, 5, 3}$, J. P. Marshall$^{3}$, F. Kemper$^{6,7,8}$, S. Srinivasan$^{9}$, \newauthor T. Dharmawardena$^{10}$, L. Fanciullo$^{11, 12, 3}$, O. Morata$^{6}$,  A. Trejo-Cruz$^{9,3}$ \\
$^{1}$Centre for Astrophysics Research, Department of Physics, Astronomy and Mathematics, College Lane Campus, University of Hertfordshire, \\Hatfield AL10 9AB, UK\\
$^{2}$Space Science Institute, 4750 Walnut Street, Suite 205, Boulder, CO 80301, USA\\
$^{3}$Institute of Astronomy and Astrophysics, Academia Sinica, 11F of AS/NTU Astronomy-Mathematics Building, No.1, Sec. 4, Roosevelt Rd, \\Taipei 106319, Taiwan\\
$^4$European Space Agency / ESTEC Keplerlaan 1, 2201 AZ, Noordwijk, The Netherlands \\
$^{5}$SRON Space Research Organisation Netherlands, Niels Bohrweg 4, 2333 CA Leiden, The Netherlands\\
$^{6}$Institut de Ci\`encies de l'Espai (ICE, CSIC), Can Magrans, s/n, E-08193 Cerdanyola del Vall\`es, Barcelona, Spain\\
$^{7}$ICREA, Pg. Llu{\'{i}}s Companys 23, E-08010 Barcelona, Spain\\
$^{8}$Institut d'Estudis Espacials de Catalunya (IEEC), E-08034 Barcelona, Spain\\
$^{9}$Instituto de Radioastronom\'ia y Astrof\'isica, UNAM. Apdo. Postal 72-3 (Xangari), Morelia, Michoac\'an 58089, Michoac\'{a}n, Mexico\\
$^{10}$Center for Cosmology and Particle Physics, Department of Physics, New York University, 726 Broadway, New York, NY 10003, USA\\
$^{11}$National Chung Hsing University, 145 Xingda Rd., South Dist., Taichung City 402, Taiwan, R.O.C.\\
$^{12}$ Tamkang University, 151 Yingzhuan Rd., Tamsui Dist., New Taipei City 251301, Taiwan, R.O.C.\\
}

\date{Accepted XXX. Received YYY; in original form ZZZ}

\pubyear{2015}

\begin{document}
\label{firstpage}
\pagerange{\pageref{firstpage}--\pageref{lastpage}}
\maketitle

\begin{abstract}
As astronomy advances and data becomes more complex, models and inference also become more expensive and complex.
In this paper we present \ampere, which aims to solve this problem using modern inference techniques such as flexible likelihood functions and likelihood-free inference.
\ampere\ can be used to do Bayesian inference even with very expensive models (hours of CPU time per model) that do not include all the features of the observations (e.g. missing lines, incomplete descriptions of PSFs, etc).
We demonstrate the power of \ampere\ using a number of simple models, including inferring the posterior mineralogy of circumstellar dust using a Monte Carlo Radiative Transfer model.
\ampere\ reproduces the input parameters well in all cases, and shows that some past studies have tended to underestimate the uncertainties that should be attached to the parameters. 
\ampere\ can be applied to a wide range of problems, and is particularly well-suited to using expensive models to interpret data.
\end{abstract}

\begin{keywords}
methods: statistical -- dust, extinction
\end{keywords}





\section{Introduction}
Modern astronomy depends heavily on interpreting data using models.
This is particularly relevant as data grows and becomes more complex, as both the volume of data to be modelled and the complexity of the models in use also grow (e.g. SKA, Vera Rubin Observatory). 
Analyses routinely involve combining diverse observations such as images, spectra, visibilities or datacubes, but even the relatively simple problem of, for example, simultaneously interpreting the global spectral energy distribution (SED) and the mid-infrared spectrum of a source is complicated by the need to interpret the continuum, lines and solid-state features in a single model.
As a result, over the last two decades the astronomical community has increasingly coalesced around Bayesian approaches as the standard for analysis \citep[see e.g.][and references therein for a useful summary, some terminology and recommendations]{Eadie_Practical_Bayesian_2023arXiv230204703E}.
Bayesian inference is an approach to inference that applies Bayes' Theorem to learn from data by encoding the information in probability distributions.
This has the advantages of properly accounting for uncertainty in data and parameters, illuminating degeneracies between model parameters, giving a principled pathway to include subjective information, and making assumptions explicit (amongst others).
However, the need to deal with probability distributions makes all but the simplest problems very computationally expensive.
As the availability of computing power grows, so too does the range of datasets and problems that Bayesian inference can be applied to, while more advanced inference algorithms become feasible.

There are also powerful reasons to take the most advanced inference approaches available.
As new facilities become more and more expensive, the \emph{cost}\footnote{where cost could be interpreted in terms of money, time, resources or energy} of observing time grows, making it ever more important to maximise the return on investment. 
This places an obligation on scientists to perform the most complete and careful analysis possible of datasets, both to maximise the output of a single observation, and to ensure that future time is not wasted on a blind alley because of incorrect inference. 
However, since powerful statistical methods have high computational (and hence energy) demands, this must be balanced against the need to use the least resources (particularly fossil fuels) possible to protect the climate \citep{Martin2022_CarbonFootprint, Knodlseser2022_CarbonFootprint}.

This use of Bayesian approaches relies on the availability of a \emph{generative} model -- that is, one that is able to explain the observations -- to solve the forward problem of inference \citep[see e.g.][]{HoggBovyLang2010}.
This forward-modelling approach is particularly important for understanding degeneracies between parameters.
In astronomy, however, a model containing all the physics that results in the observable may be impractical or impossible, either because there is no complete solution or because it requires too much computing time to reach it.
If a model is incomplete in such a way that it fails to produce key parts of the observed data (model misspecification), it may bias the inference.
For example, if emission lines are missing from a model of a spectrum, the continuum may then be overestimated to compensate, with knock-on effects to other parameters of interest.

In recent years, there have been several developments which help to alleviate model misspecification.
In particular, the development of \emph{flexible likelihood functions} \citep[see e.g.][for an application to astronomy]{Starfish_paper}, which enable self-consistent down-weighting of features that the model is incapable of explaining.
This simultaneously enables the best inference possible with incomplete models, while also identifying how to improve future models.
This in turn reduces the required computing power without biasing inference and hence wasting effort.

Nevertheless, many problems can only be solved with models that are still numerically intensive, even in an incomplete form.
For example, understanding the mineralogy of newly-formed dust in a variety of environments relies on solving the radiative-transfer problem to predict the emission of the dust, particularly in the mid-infrared regime where many dust species have features.
The shape and strength of these features are influenced by e.g. the size, shape, stoichiometry, or mineralogy of the dust grains, as well as radiative-transfer effects, making this a highly-degenerate problem \citep[e.g.][]{Srinivasan2017}; as a result, many studies make large numbers of simplifying assumptions to reduce the size of the parameter space, and hence fail to quantify the impact of these degeneracies on their results.
Nevertheless, even an incomplete solution to the radiative-transfer problem (e.g. considering stars and dust only, not the coupling with gas, and treating the geometry in a reduced symmetry such as 1D or 2D instead of arbitrary dust distributions) may take up to several minutes even with GPU acceleration \citep[e.g.][]{Juvela2019_SOC}.
Similarly, computing the cross-sections of a distribution of dust grains from their refractive indices alone takes seconds even in simplified cases such as Mie theory \citep[compact spherical grains][]{Mie1908}.
In these cases, further optimisation of the inference is required beyond that which can be achieved with flexible likelihood functions.
Some works have opted to accelerate this by using neural networks to train fast surrogate models on a small number of simulations of the exact model \citep[e.g.][]{Ribas2020} and then deploying this approximate model with Markov Chain Monte Carlo (MCMC) for ``exact'' inference.
However, in fields such as cosmology there is a long history of instead continuing to use the exact but expensive model and adopting approximate inference methods like Approximate Bayesian Computing \citep[e.g.][and citations therein]{Beaumont_abc_review, Grazian_abc_review}.
In recent years there has been particular growth in the use of neural estimation methods, which exploit a neural network to directly infer the likelihood, posterior or likelihood-to-evidence ratio from simulations of the observations (see e.g. \citealt{Cranmer_sbi_review} for a recent review and \citealt{Hahn_ANPE_2022ApJ...938...11H} for a recent example in astronomy). 


In this paper, we present a new software package, \ampere, to handle the issues identified in the preceding paragraphs. \ampere\ is specifically designed to handle automatic inference on astronomical data with misspecified and/or very expensive models. In Sect.~\ref{sec:amp} we discuss the structure and capabilities of \ampere, which are then demonstrated on synthetic observables and literature datasets in Sect.~\ref{sec:syn}. Finally, we summarise the manuscript in Sect.~\ref{sec:conc}.


\section{\ampere}\label{sec:amp}
\ampere, available at \url{https://github.com/ICSM/ampere.git}, is a framework for automatic inference on incomplete astrophysical models.
It prioritises comparing physical models to astronomical observations e.g. spectra, but can in principle be adapted to a wide range of problems.
By providing a means for users to define their model in terms of a data-generating process and a prior on its parameters, \ampere\ aims to make Bayesian inference easy even for complex datasets.
In this section, we provide a description of the key approaches underlying \ampere, along with a description of the code, its capabilities and features.

\subsection{Flexible likelihood functions for model misspecification}

\ampere\ employs a flexible likelihood approach \citep[e.g.][]{Starfish_paper} to handle imperfect forward models by treating structured residuals as arising from correlated noise in the data.

\begin{figure*}
    \includegraphics[width=\textwidth]{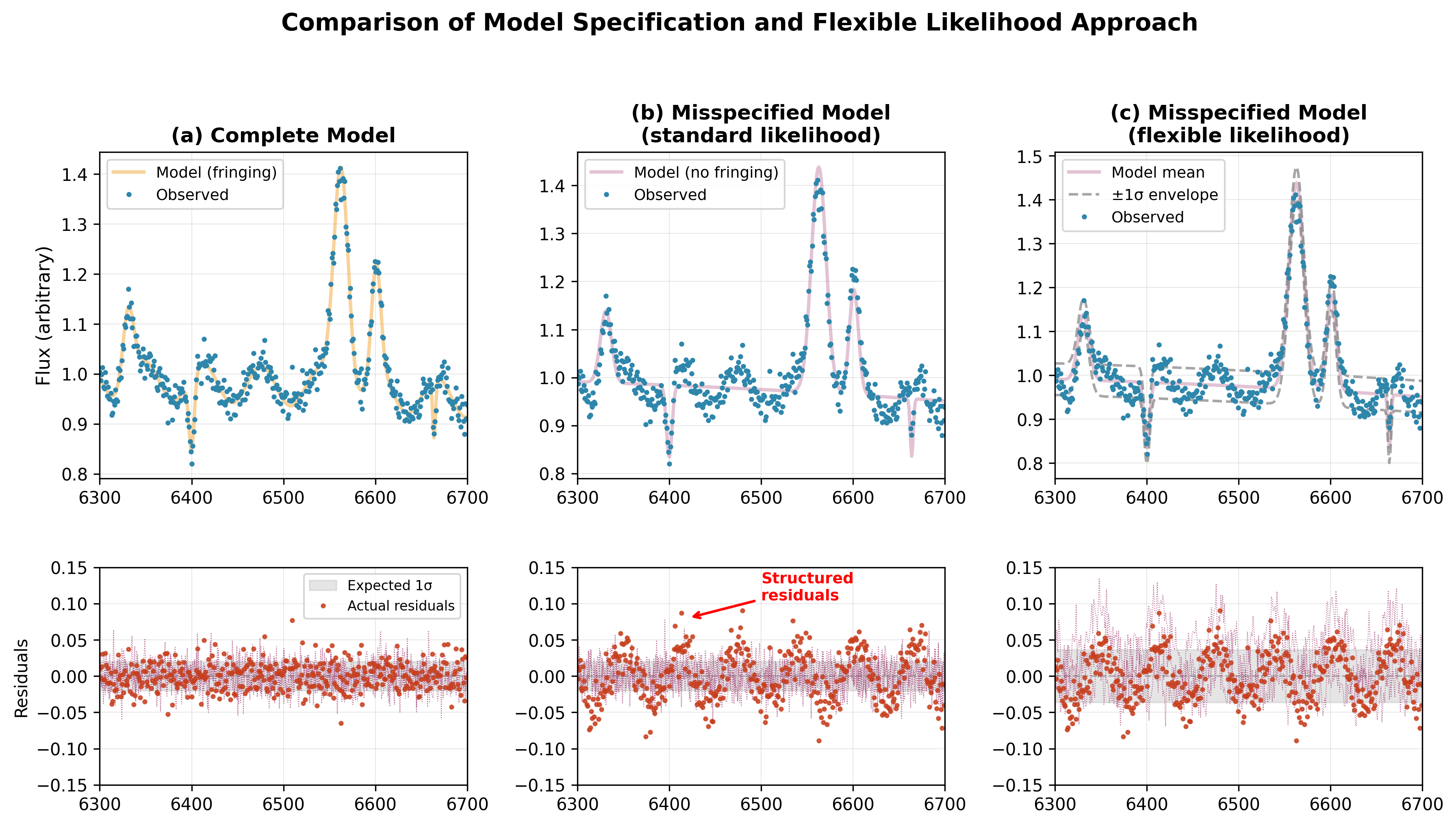}
    \caption{Demonstration of concepts of model misspecification and flexible likelihoods. If we start from a spectrum that contains emission and absorption lines but is contaminated by fringing, we can attempt to interpret it by modelling the lines and the fringing simultaneously (\emph{a, top}). When using this model, the residuals (red points, \emph{a, bottom}) follow the expected i.i.d. Gaussian distribution (three samples from which are shown as pink dotted lines). However, if we take a misspecified model which does not include the fringing (\emph{b, top}), the resulting residuals (\emph{b, bottom}) show quasi-periodic structure, which is inconsistent with the samples from the i.i.d. Gaussians we would normally expect (pink dotted lines), resulting in a dramatic misunderstanding of the uncertainty and potentially the parameters as well. However, when we change the likelihood function (\emph{c, top}), the uncertainty distribution becomes more representative. Indeed, the simplified uncertainty band is overly pessimistic, since it neglects the dependence between points that has been introduced (\emph{c, bottom}. The samples from the correlated distribution in this case show not just similar amplitude to the observed residuals, but also oscillations of similar quasi-period and amplitude. This highlights the equivalence between model misspecification and correlated noise. }\label{fig:flexlikeexample}
\end{figure*}

Model misspecification produces structured residuals that are statistically indistinguishable from correlated observational noise \citep[e.g. Fig. 5 in ][]{Starfish_paper}. Consider the case of a spectrum with fringing (see Fig.~\ref{fig:flexlikeexample}). The fringes themselves can be considered a physical, quasi-periodic component of the model that has to be introduced, or a source of quasi-periodic correlated noise. However, from the perspective of the residuals produced by a model that does not consider fringing, either of these possibilities are equivalent.

Since these two effects are indistinguishable, we can therefore absorb model deficiencies into the data covariance matrix. Given observations $d$ (which could be a spectrum, SED, or other observable, with length $n$)\footnote{This section assumes the data is a vector, but the problem can be trivially extended to higher-order cases e.g. images, multiple vectors, etc.} with nominal uncertainties $\vec{\sigma}$, we assume that the data are drawn from a multivariate Normal distribution such that
\begin{equation}
    d \sim \mathcal{N}\left(m(\vec{\theta}), \mathbf{C}( \vec{\phi})\right),  
\end{equation}
where $m(\vec{\theta})$ is the model prediction. Note that this prescription naturally separates the physical parameters $\vec{\theta}$, and the parameters of the noise model $\vec{\phi}$. 
This approach arises from the assumption that the uncorrelated noise is drawn from a multivariate Normal such that $d \sim \mathcal{N}\left(m(\vec{\theta}), \sigma^2\right)$ and that either correlated noise or model misspecification must be incorporated. However, it does not distinguish between these two potential causes of structure.

The covariance matrix decomposes as
\begin{equation}
    C_{ij}(\vec{\phi}) = \mathcal{K}(\lambda_i, \lambda_j; \vec{\phi}) + \delta_{ij}\ \sigma^{2}_{i},
\end{equation}
where 
$\delta_{ij}$ is the Kronecker delta,
$\lambda_i$ is the wavelength of the $i$-th data point, $\sigma_i$ are the known observational uncertainties, and $\mathcal{K}$ is a kernel function encoding correlations. Multiple 
effects may be represented by adding kernels:
\begin{equation}
    \mathcal{K}(\lambda_i, \lambda_j; \vec{\phi}) = \sum_{\ell=1}^{L} \mathcal{K}_\ell(\lambda_i, \lambda_j; \vec{\phi}_\ell).
\end{equation}
This construction ensures $\mathbf{C}$ remains positive (semi-)definite while limiting the dimensionality of $\vec{\phi}$ \citep[e.g.,][]{rasmussen2006gaussian}.


The resulting log-likelihood is
\begin{multline}
    \ln \mathcal{L}(\mathbf{d} \mid \vec{\theta}, \vec{\phi}) = -\frac{1}{2}\left[n\ln(2\pi) + \ln|\mathbf{C}(\vec{\phi})| \right. \\
    + \left. r(\vec{\theta})^T\mathbf{C}(\vec{\phi})^{-1}r(\vec{\theta})\right], \label{eq:loglike}
\end{multline}
where $r = d - m(\vec{\theta})$ are the residuals. 
The astute observer will notice that this is equivalent to the log marginal likelihood of a Gaussian process whose mean function is the physical model i,e, $m(\vec{\theta})$, kernel $\mathcal{K}$ and hyperparameters $\vec{\phi}$. 

Unconstrained maximization over $\vec{\phi}$ would allow us to trivially inflate $\mathcal{K}$ to allow for any residual structure. It therefore behoves us to impose priors $p(\vec{\phi})$ on the hyperparameters and marginalize:
\begin{equation}
    p(\vec{\theta} \mid \mathbf{d}) \propto p(\vec{\theta}) \int p(\mathbf{d} \mid \vec{\theta}, \vec{\phi}) \, p(\vec{\phi}) \, \mathrm{d}\vec{\phi}.
\end{equation}
By marginalising over these hyperparameters, we ensure that the distribution of correlated noise is accounted for without allowing the noise to grow arbitrarily large. 
This in turn automatically implies that we should perform Bayesian inference over $\vec{\theta}$ since we are already paying the computational cost involved.


\subsection{\ampere\ architecture \& features}
\begin{figure}
    \includegraphics[width=\columnwidth]{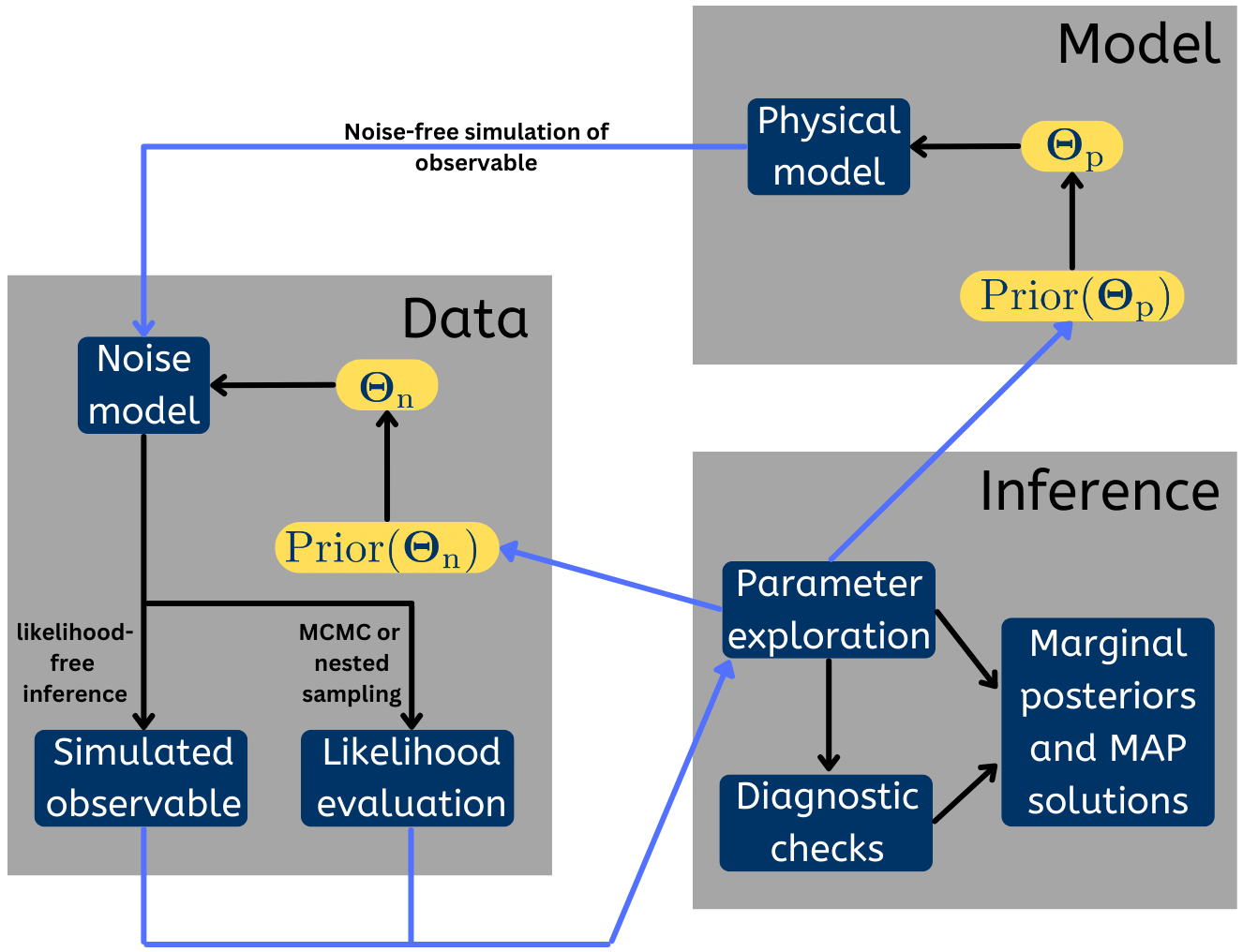}
    \caption{\ampere\ architecture. See text for details.}
    \label{fig:architecture}
\end{figure}

\ampere\ is designed to be modular and extensible, and hence employs object-oriented approaches throughout the code. 
It defines three types of objects: 1) model; 2) data; and 3) inference objects (Figure \ref{fig:architecture}).
Through them, \ampere\ defines a uniform interface to perform inference on arbitrary (user-defined) models.
This allows users to solve their own problems while making it easy to extend and maintain the codebase.
The following sections expound further on the implementation and usage of these objects.

\subsubsection{Models}
Model classes define a physical model ($M$ above) for which the user wishes to perform inference.
The model packages the metadata along with a method which forward-models (i.e. simulates) realisations of the model free from observational effects (e.g. noise, PSFs, etc.).
Models must also provide methods to compute the prior probability density of a set of model parameters, and a method to transform from samples drawn from the Uniform distribution to those drawn from the prior. The choice of inference method may place constraints on how the prior is defined.
Models may be deterministic or stochastic, but care must be taken with stochastic models - not all inference methods are appropriate in these cases.
While a number of basic models are included with \ampere\footnote{including all models used to demonstrate and validate \ampere\ in this work}, it is generally assumed that users will define their own models suitable to their problem.

\subsubsection{Data}
Data classes encapsulate the processes required to go from the physical model to the observables, i.e. they contain the noise model, including any additional correlated noise as described above, as well as any transformations that the instrument applies to the data. 
At present, \ampere\ implements classes to handle photometric and spectroscopic observations.
Processes implemented in data classes might include convolution with a point-spread function or line-spread function, convolution with a filter profile, resampling in wavelength or angular coordinates, etc.
It should also include effects such as zero-point offsets between different observations (e.g. spectral orders).
All of these effects should either be precisely known, or treated as nuisance parameters to be integrated out.
Finally, data objects evaluate the log-likelihood of the data given an evaluation of the model (for likelihood-based inference), or generate sample (i.e. noisy) observables given a model realisation (for simulation-based, or likelihood-free, inference).

Photometry are assumed to have uncertainties supplied by the user that include any e.g. calibration uncertainty, and that the data points are uncorrelated. 
However, the user is free to supply a fixed covariance matrix if the instruments are sufficiently well characterised.
The only transformation applied is therefore to convolve the model spectrum with filter profiles to obtain appropriate model flux estimates.
Synthetic photometry is implemented using \emph{pyphot} \citep{pyphot}.

Spectra, on the other hand, are treated with a slightly more complex noise model. 
The uncertainties on the data are assumed to be uncorrelated, and include only the relative uncertainty between points, i.e. no calibration uncertainty.
The user is therefore able to supply a prior on the calibration (assumed to be drawn from a log-normal with $\mu=0, \sigma=0.01$ otherwise), and the observed spectra are multiplied by this factor before comparing to the model.
As a result, each order of a spectrum should be treated as a separate observation to optimise their combination and correctly propagate the stitching uncertainty into the model results.
Spectrum objects also implement the flexible likelihood function described above. 
By default, this features a single Gaussian kernel (also known as a radial basis function or RBF kernel) per spectrum object to account for additional correlated noise. 
Future improvements will enable a broader range of kernels, including user-specified ones.

Future development will implement data classes to handle a wider range of observations, such as (spectro-)imaging, polarimetry, or interferometric observables.
Furthermore, we expect to implement higher-level data containers (e.g. datasets, samples) to facilitate hierarchical inference.

\subsubsection{Inference}
\ampere\ uses inference objects to provide an interface to a wide range of packages for Bayesian inference.
Rather than reinventing the wheel and performing inference ourselves, we wrap other packages and ensure that they receive the information they expect.
At present, \ampere\ supports inference with MCMC using \emph{emcee} \citep{emcee_paper} and \emph{zeus} \citep{karamanis2020ensemble, karamanis2021zeus}, as well as nested sampling with \emph{dynesty} \citep{dynesty_v1_paper}. 
These methods all rely on directly evaluating the likelihood, and so may give strange behaviour for stochastic models\footnote{The stochasticity of the model means that any single likelihood evaluation is only one of a distribution for that set of parameters. It is therefore necessary to estimate a marginal log-likelihood for stochastic models by computing multiple realisations of the model. While this will be implemented in \ampere\ in future, it is rather inefficient compared to approaches which are designed for stochastic inference.}.
We have therefore also implemented likelihood-free inference using \emph{sbi} \citep[Simulation-based inference;][]{tejero-cantero2020sbi}; \emph{sbi} provides access to a number of different likelihood-free approaches which are often orders of magnitude more efficient for models which are stochastic or have very high computational cost. 
\ampere\ provides an interface to the ``\emph{(Sequential) Neural Posterior Estimation}'' \citep[(S)NPE,;][]{NPE_NIPS2016_6aca9700} approach as implemented in \emph{sbi}, which uses a masked autoregressive flow neural network to directly learn the relationship between the parameters of the model (including noise) and observations.
In particular, NPE provides access to \emph{amortized inference}, in which an approximate posterior can be learned from a set of simulated observations and then applied to many observations without re-training, providing almost instantaneous inference for any number of sources \citep{amortised_neural_inference}.
Through these 4 packages, \ampere\ provides tools suitable for a wide range of astronomical inference problems.
The characteristics of the different approaches are summarised in Table~\ref{tab:infcodes}.

\begin{table*}
\caption{Comparison of currently-implemented inference approaches in \ampere}   
    \label{tab:infcodes}
    \centering
    \begin{tabular}{c|cccc}
        \multirow{2}{*}{}{approach} & \multirow{2}{*}{}{\shortstack{affine-invariant\\ MCMC}} & \multirow{2}{*}{}{\shortstack{slice-sampling\\MCMC}} & \multirow{2}{*}{}{\shortstack{nested\\ sampling}} & \multirow{2}{*}{}{\shortstack{likelihood-free\\inference}}\\\hline
        package  & emcee & zeus & dynesty & sbi \\
        max \# parameters  & few tens & few hundred & few hundred & depends$^{\ast}$ \\
        correlated parameters & \checkmark & \checkmark & \checkmark & \checkmark \\
        multimodal posteriors & \text{\sffamily X} & \checkmark & \checkmark & \checkmark \\
        requires gradients & \text{\sffamily X} & \text{\sffamily X} & \text{\sffamily X} & \text{\sffamily X} \\
        stochastic models & \text{\sffamily X}  & \text{\sffamily X} & \text{\sffamily X} & \checkmark \\
        amortized inference & \text{\sffamily X} & \text{\sffamily X}& \text{\sffamily X}& \checkmark \\
        \hline
        \multicolumn{5}{l}{$^{\ast}$The LFI approaches used in \ampere\ are insensitive to nuisance parameters - the algorithm}\\ \multicolumn{5}{l}{simply ``sees'' them as a source of noise, which therefore do not impact the scaling of the inference.}
    \end{tabular}
\end{table*}

Inference objects also provide post-processing capabilities.
\ampere\ automatically computes some statistics commonly used to identify non-convergence in MCMC samplers, such as the effective sample size (ESS), autocorrelation, Rubin-Gelman split-$\mathrm{\hat{R}}$ \citep{Rhat1992} and Geweke-$z$ drift score \citep{geweke}.
\ampere\ will issue warnings when these statistics clearly indicate non-convergence.
\ampere\ also produces a number of diagnostic plots, including posterior-predictive plots, traces, and corner plots.
Some examples of these can be seen in the following sections.


\section{Practical applications of \ampere}\label{sec:syn}
What kind of thing can \ampere\ do? In this section, we demonstrate \ampere\ with a number of test models, demonstrating that it recovers the input parameters for a small selection of common astronomical inference problems. Code to reproduce all examples is available at \url{https://github.com/ICSM/ampere/tree/master/examples}.


\subsection{A modified blackbody}\label{sec:mbb}
The modified blackbody (or Planck) function is one of the most widely applied in astronomy. 
It is particularly useful for inferring the temperature and emissivity index of cold, featureless dust.
However, it suffers from a number of issues that require care: the model is not linear in its parameters, making many simplifying assumptions useless, and there is a degeneracy between two of the parameters, the temperature $T$ and the spectral index $\beta$.
This degeneracy makes it imperative to understand the posterior distributions of the parameters under this model.

As a simple test, we therefore define an arbitrary \emph{true} modified blackbody with parameters ($T, \beta, \log_{10} \frac{M}{\rm M_\odot}, d$) = (30\,K, -2, 1, 0.11\,kpc) where $M$ is the dust mass and $d$ is the distance to the object, assuming a fixed mass-absorption coefficient at the wavelength 250\,$\mu$m $\kappa_{250} = $10\,cm$^{2}$\,g$^{-1}$. From the true model, we generate synthetic photometry in the AKARI FIS and \emph{Herschel} PACS~\&~SPIRE bands (i.e. ten bands in total), and add Gaussian noise to them assuming an uncertainty of 10\% of the true flux value and no correlated noise. We adopt independent priors over each parameter of $P\left(T\right) \sim U\left(10, 50\right)$, $P\left(\beta\right)\sim U\left(-3, 0\right)$, $P\left(\log\frac{M}{\rm M_\odot}\right)\sim U\left(-3, 3\right)$ and $P\left(d\right) \sim U\left(0.05, 0.15\right)$ and retrieve the known input parameters using each inference method implemented in ampere.

\begin{figure*}
    \centering
    \includegraphics[width=\textwidth]{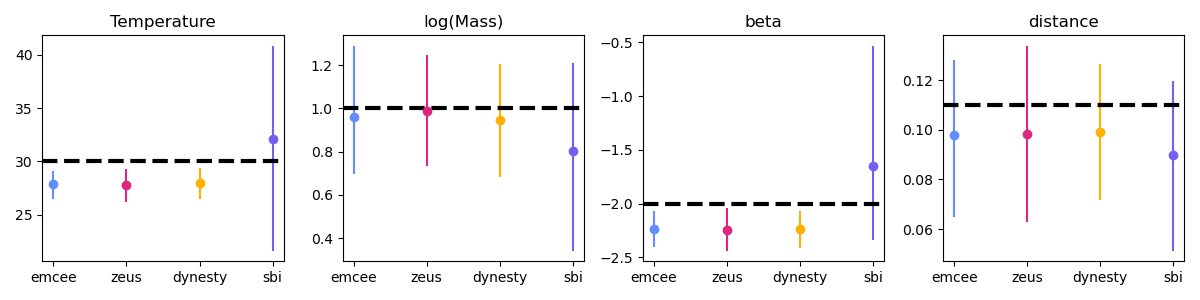}
    \caption{Composite figure showing the values recovered by the different inference approaches for the modified blackbody example. The black dashed line in each panel denotes the 'true' values of the input blackbody function, whilst the coloured dots denote the output for the four inference methods implemented in \ampere.}
    \label{fig:ex_mbb}
\end{figure*}

As seen in fig.~\ref{fig:ex_mbb}, all four approaches recover the input parameters.
All models were run for approximately the same amount of wall time on the same machine, and made similar estimates of the median of the posterior 1-D marginal distributions of the parameters.
However, \emph{sbi} produced a larger estimate of the credible interval, as expected for an approximate inference method.
This typically occurs when there are not enough sample models to accurately describe the distribution of the data.
This could be remedied by producing more sample models or by instead using multiple rounds of sampling.
However, given the simple nature of this model and data, \emph{sbi} is unnecessary for this case, and is only used here for demonstration.

\subsection{Astrophysical parameters of stars}\label{sec:aps}
A second key problem in astronomy is to infer the properties of stars. 
In the age of massive photometric and spectroscopic surveys, doing this \emph{quickly and efficiently} is doubly important.
The properties of stars are used to understand the structure, composition and history of galaxies (particularly our own).

To demonstrate this problem, we create a synthetic dataset and attempt to recover the input parameters with \ampere. 
The star is generated from PHOENIX model spectra for a given ($L, T_{\rm eff}, \log g)$ combination, and foreground extinction ($R_{\rm V}, A_{\rm V}$)
The parameter recovery is tested with both a typical set of photometric filters alone (Gaia: G, BP \& RP; 2MASS: JHK$_{s}$; WISE: W1 \& W2), and in combination with a synthetic spectrum similar to a typical \emph{Gaia} RVS spectrum (842 nm $\leq \lambda \leq$ 872 nm; $R\sim 11000; {\rm SNR}\sim100$).
The model then uses {\sc starfish} \citep{Starfish_paper} to interpolate in the space of model spectra and {\sc extinction} \citep{Extinctionpkg} to compute foreground extinction.
Inference on this model is performed with sbi and 10,000 model evaluations and the default masked autoregressive flow posterior for NPE.

\begin{figure*}
    \centering
    \includegraphics[width=0.85\textwidth]{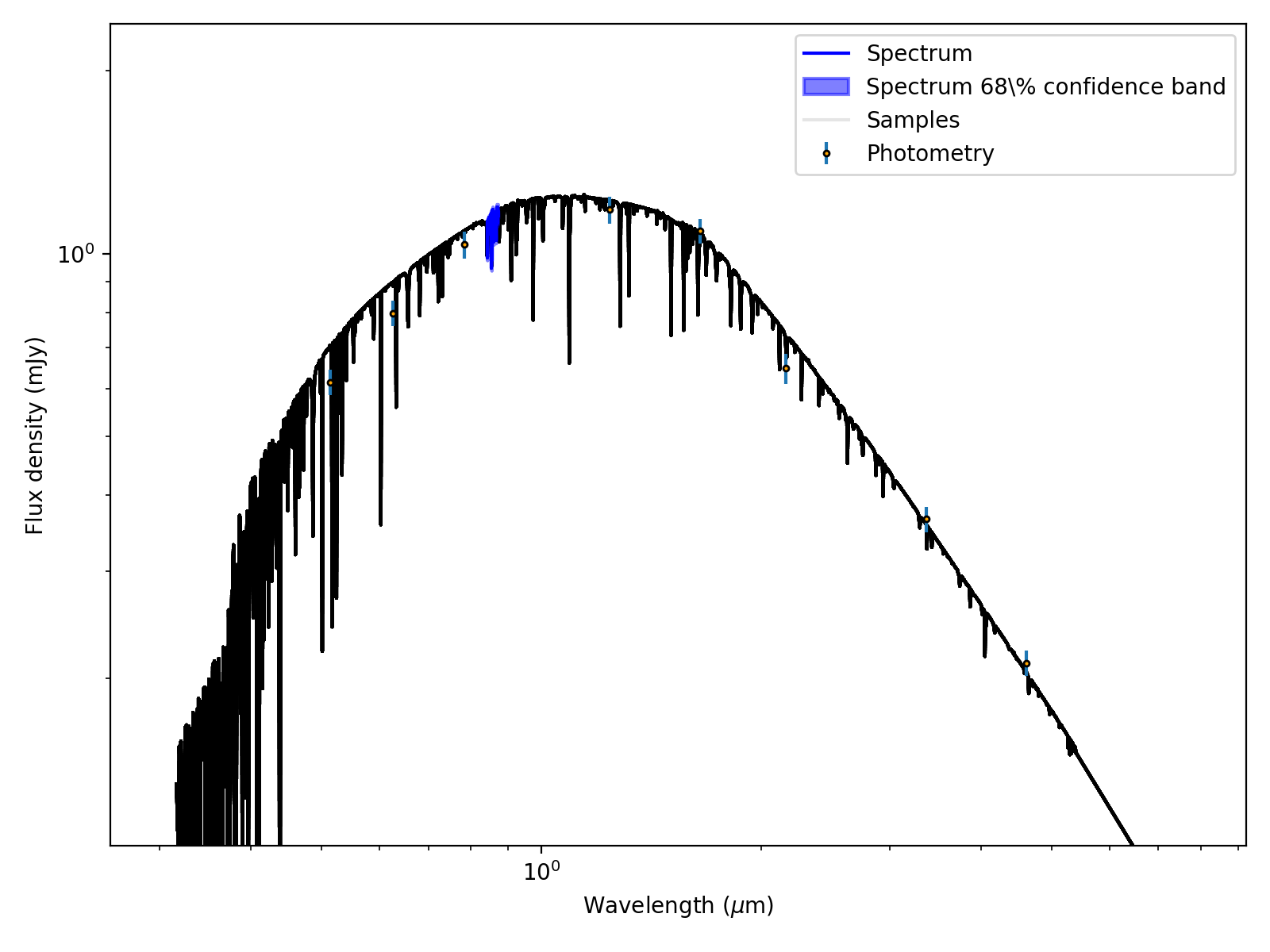}
    \caption{Posterior samples from models fitted to synthetic photometry and a synthetic Gaia spectrum for inferring astrophysical parameters. The parameters are so tightly constrained by this combination that the model samples form a single band.}
    \label{fig:photovspec}
\end{figure*}

\begin{table}
    \centering
    \caption{Results of fitting synthetic photometric and spectroscopic observations with \ampere}
    \label{tab:APrecovery}
    \begin{tabular}{lccc}
        Parameter & Truth & Phot & Spec  \\\hline
        $\log\left(L\right)$ & 0.67&$0.68\pm0.10$&$0.61\pm0.02$\\
        $T_{\rm eff}$ &6750&$6940^{+600}_{-540}$&$6527^{+49}_{-50}$\\
        $\log\left(g\right)$&4.37& $4.5\pm0.3$&$4.1\pm0.1$\\
        $A_{\rm V}$&1.0&$1.0\pm0.3$&$0.81\pm0.07$\\
        $R_{\rm V}$&3.2&$3.4^{+0.4}_{-0.3}$&$3.44^{+0.29}_{-0.28}$
    \end{tabular}
\end{table}

The results are shown in Fig.~\ref{fig:corneraps} and Table~\ref{tab:APrecovery}. 
We see that \ampere\, is highly effective when working with either photometry or both photometry and spectroscopy.
It efficiently recovers the input parameters, albeit with a small bias in the case of spectroscopy. 
The final parameter uncertainties are comparable to published values from \textit{Gaia} RVS spectra \citep[e.g.][]{Recio-Blanco2023A&A...674A..29R}. 
Given that it took approximately 20 minutes to generate the models and train this amortised posterior, this shows that the adoption of amortised approaches can be used to dramatically accelerate inference for large photometric and spectroscopic surveys without loss of precision.

\subsection{A star with circumstellar dust}\label{sec:starplusdust}

Now, we consider a combination of the above two problems - a star orbited by dust whose emission follows a modified blackbody. 
This is analogous to the detection and characterisation of debris disc host stars, or other cases of stars with cold, featureless circumstellar dust.
In this particular case, we attempt to fit observations of HD~105, a young Sun-like star surrounded by a circumstellar debris disc \citep{2018Marshall}.
We model the stellar photosphere following the method laid out in Sect.~\ref{sec:aps}, and include the dust emission as a modified blackbody.
Since the dust component radiates at much longer wavelengths than the star, we add longer wavelength synthetic photometry to the model in the \emph{Herschel}/PACS and SPIRE filters, although we omit the ALMA and ATCA photometry included in \cite{2018Marshall} from the modelling presented here.

The parameters of the model are therefore $L_{\star}$ (in $L_{\odot}$), $T_{\star}$ (in K), $\log g$,  and metallicity [Fe/H] (in dex, relative to Solar) for the star, and the dust temperature, $T_{\rm dust}$ (in K), emitting area $A_{\rm dust}$ (in au$^{2}$), break wavelength $\lambda_{\rm 0}$ (in $\mu$m) where the slope of the dust emissitivity changes and spectral index $\beta$ for the dust component.
We adopt uniform priors for all parameters of the model, restricting the ranges to physically meaningful values.
For the stellar parameters, we first fixed the distance for the system to that inferred from the Gaia DR3 parallax measurement. 
The constraint on the distance strongly curtails the set of $L_{\star}$, $T_{\star}$ values that can then fit the photometry, reducing the impact of the inherent degeneracy of those parameters.
The stellar luminosity was constrained within a range 1.0 to 1.4 $L_{\odot}$, and the stellar temperature within a range 5,500 to 6,500~K, bracketing the expected values of 1.2~$L_{\odot}$ and 6034~K.
The log stellar surface gravity was given a range of 4.0 to 5.0, and finally the stellar metallicity was constrained between 0.0 and 0.5 $Z_{\odot}$, again with expected values of 4.478 and 0.02, respectively.
For the dust parameters, we restrict the range of $T_{\rm dust}$ between 30 to 100~K, the area scaling factor $A_{\rm dust}$ has an allowed range between $10^{-3}$ and $10^{3}$ au$^{2}$, the break wavelength $\lambda_{\rm 0}$ is restricted to between 50 and 500~$\mu$m, and the power law index $\beta$ value is assumed to lie between 0 and 4. The ranges of these parameters are motivated by either the observational constraints, or the known properties of debris discs \citep[e.g.][]{1993prpl.conf.1253B,2008ARA&A..46..339W}. Due to its large dynamic range, the scaling factor $A_{\rm dust}$ is explored in log-space to help with convergence.

Fitting was carried out using emcee to explore parameter space using 40 walkers over 10,000 steps to generate 400,000 realisations of the model. A burn-in of 160,000 realisations was discarded before construction of the posterior probability distribution and estimation of parameter values and uncertainties from the remaining 240,000 realisations. 
Realisations of the star plus disc model were generated by \ampere\ using the QuickSED script held under the directory: {\tt ./ampere/models/QuickSED.py}.
The modelling results are held in the \ampere\ repository under the directory {\tt ./examples/star\_disc/} with a timestamp: {\tt 2023-08-18\_11-07-25}. Total run time for \ampere\ to produce this example was $\simeq$27 hours using 32 Intel Xeon 2.1 GHz processors, including the downloading and training of the {\sc Starfish} grid of model stellar atmospheres.

\begin{figure*}
    \centering
    \includegraphics[width=0.85\textwidth]{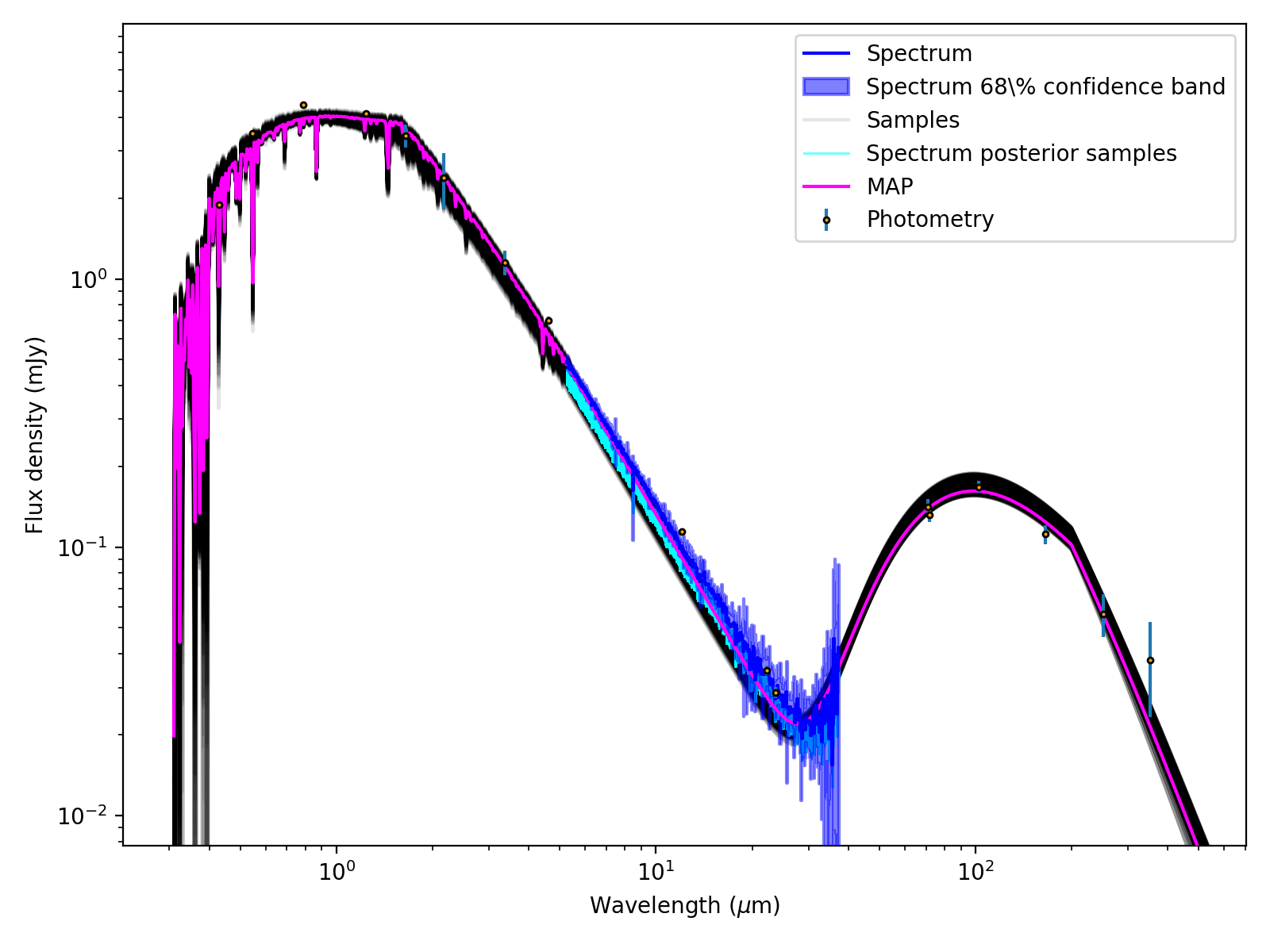}
    \caption{Plot of wavelength vs flux density for the star with circumstellar dust example. Photometric data points are denoted by the circular data points with 1-$\sigma$ uncertainties, the \textit{Spitzer}/IRS spectrum is denoted by the blue line and its 1-$\sigma$ uncertainty as the shaded region. Realisations of the stellar photosphere and dust modified blackbody model are shown by the black lines, with the maximum amplitude probability model denoted by the magenta line.}
    \label{fig:starndust_posterior}
\end{figure*}

The results of the fitting process reproduce the observations well.
The maximum \emph{a posteriori} parameters for the system are broadly consistent with the values presented in \cite{2018Marshall}, where they can be directly compared with their visualisation in Figure \ref{fig:starndust_posterior}. 
The posterior probability distributions for the model run included with the repository is shown in Figure \ref{fig:starndust_corner}, and the numerical results of the fitting are presented in Table \ref{tab:starndust}.
The clear exception to the general consistency between these approaches is the $\beta$ value in the \ampere\ fit being significantly steeper than the fit determined in the original paper, but that may be attributed to the absence of millimetre wavelength data points.

\begin{table}
    \centering
    \caption{Results of star with circumstellar dust.}
    \label{tab:starndust}
    \begin{tabular}{lcc}
        Parameter & \citet{2018Marshall} & This work \\
        \hline
        $L_{\star}$ ($L_{\odot}$) & 1.216$\pm$0.005 & 1.298$^{+0.001}_{-0.001}$ \\
        $T_{\star}$ (K) & 6034$\pm$8 & 5784$^{+0.2}_{-0.2}$ \\
        $\log g$ & 4.478$\pm$0.006 & 4.32$^{+0.03}_{0.04}$ \\
        $\left[\mathrm{Fe/H}\right]$ (dex) & 0.02$\pm$0.04 & 0.07$^{+0.02}_{-0.01}$ \\
        $A_{\rm dust}$ (au$^{2}$) & 0.15 & 0.16$^{+0.21}_{-0.10}$ \\
        $T_{\rm dust}$ (K) & 50.0 & 51.3$\pm$0.2 \\
        $\lambda_{\rm 0}$ ($\mu$m) & 100.0 & 99$^{+0.1}_{-0.1}$ \\
        $\beta$ & 1.0 & 1.37$\pm$0.06 \\
    \end{tabular}
\end{table}

\subsection{Mineralogy of LMC carbon-star envelopes}\label{sec:cstar}
Following from above, we expect an effective use-case for \ampere\ to be determining mineralogy of dusty environments, and we illustrate this by examination of the modelling of an asymptotic giant branch star.
The outflow of the LMC carbon star OGLE LMC LPV 28579 was modelled in detail by \citet{Srinivasan2010}, however, they performed the inference by-eye to determine the optimal abundance ratio of amorphous carbon to silicon carbide (SiC) dust.
Here we repeat their analysis using \ampere\ to perform inference instead.
Since this is only a demonstration, we fix all the model parameters except the relative abundances of the dust, the stellar luminosity, and inner and outer radii of the dust shell. There are also two nuisance parameters, the stellar mass and $r_0$ for the power-law density distribution which are sampled but do not contribute to the output spectrum. For the fixed parameters (e.g. the parameters of the stellar photosphere), we adopt the best-fitting parameters found by \citet{Srinivasan2010}. 
We stress that this choice is simply made to make the model presentation easier, and there is no reason why all parameters could not be inferred simultaneously.
We use the Monte Carlo Radiative Transfer code {\sc hyperion} \citep{Hyperion_paper} for the dust emission calculations.
This was a somewhat arbitrary choice, as any appropriate radiative transfer code that can be called in a scripted fashion would be appropriate using the modular nature of \ampere. 
Since this model relies on calculating dust radiative-transfer, we use NPE to keep the inference time reasonable;
as this is necessarily a stochastic model NPE is better able to handle the inference than \ampere's other options.

We use the same photometry and spectra presented in \citet{Srinivasan2010}, and perform NPE using a masked autoregressive flow density estimator. 
We use 10\,000 model evaluations to train the posterior.
The priors on all parameters are uniform, except for the relative abundances which are drawn from a Dirichlet distribution with concentration parameter $\alpha = 1$. 
The results of our fit are shown in Table~\ref{tab:agb}, alongside the results from \citet{Srinivasan2010}; the 1D and 2D marginal posterior distributions are shown in Fig.~\ref{fig:cstar-corner}.
We see that our results are in general consistent with those of \citet{Srinivasan2010}, however our uncertainties suggest that the approach of \citet{Srinivasan2010} tended to overestimate the precision of their results by a factor of a few.
One discrepancy is seen in the relative abundance of SiC and amorphous carbon.
This is probably caused by the use of \citet{Rouleau1991ApJ...377..526R} optical constants here, which produce dust grains a few times less emissive than the \citet{Zubko1996MNRAS.282.1321Z} optical constants used in \citet{Srinivasan2010}; hence, a larger fraction of amorphous carbon is required to maintain the same feature-to-continuum ratio.
We also see a strong correlation between R$_{\rm out}$ and M$_{\rm d}$; this is to be expected, given that increasing the envelope size increases the mass of the shell, but it bears emphasising that \emph{the uncertainty on the outer radius is the main driver of uncertainty in the dust mass}.

Generating these results with sbi takes only a matter of days (most of which is spent generating the radiative-transfer models), while experiments with MCMC took many weeks merely for burn-in.
If this were used for amortised inference, the investment in generating models and training could be used to generate posterior samples for huge samples of AGB stars in a fraction of the time required for MCMC to sample for even a single source.

\begingroup
\renewcommand{\arraystretch}{1.3}
\begin{table}
    \centering
    \caption{OGLE LMC LPV 28579 results}
    \label{tab:agb}
    \begin{tabular}{lcc}
         Parameter & \citet{Srinivasan2010} & This work\\\hline
         L$_\ast$ [L$_\odot$]&4810 & $4909^{+240}_{-244}$ \\
         $\log \frac{\mathrm{R}_{\rm in}}{\mathrm{R}_{\ast}}$ &0.64&$0.34\pm1.15$\\
         $\log \frac{\mathrm{R}_{\rm out}}{\mathrm{R}_{\rm in}}$ &3& $2.82\pm0.35$\\
         $\log \frac{\mathrm{M}_d}{\mathrm{M}_\odot}$ &$-5.85^\dagger$& $-6.02\pm0.21$\\
         $\log \frac{\dot{\rm M}_d}{\mathrm{M}_\odot \mathrm{yr}^{-1}}$ &$-8.60^{+0.06}_{-0.02}$&$-8.60^\dagger$ \\
         C fraction &$0.88^{+0.02}_{-0.04} $& $0.949\pm0.056$\\
         SiC fraction &$0.12^{+0.04}_{-0.02} $&$0.045^{+0.043}_{-0.042}$\\
         \hline
         \multicolumn{3}{l}{$^\dagger$Converted from the point estimate using $\dot{M}_{\rm d} = \frac{M_d}{\frac{R_{\rm out}}{R_{\rm in}} R_{\rm in}} v_{\rm exp}$, where }\\
         \multicolumn{3}{l}{we assume $v_{\rm exp} = 10$\,km\,s$^{-1}$.}
    \end{tabular}

\end{table}
\endgroup


\subsection{The detection of carbonates in NGC 6302}\label{sec:ngc6302}
In 2002, \citeauthor{Kemper_02_Detection} discovered carbonates outside the Solar System in planetary nebula NGC 6302. 
Carbonates are normally associated with aqueous alteration (weathering) of silicates (rocks), which requires the presence of liquid water on a parent body with an atmosphere. 
Due to the partial pressure from the atmosphere, gaseous CO$_2$ is dissolved in the water as CO$_3$$^{2-}$, which can then react with the rocky material to form carbonates such as calcite (CaCO$_3$) and dolomite (CaMg(CO$_3$)$_2$) (the species detected in NGC 6302). 
The amount of carbonates detected in NGC 6302 is significant, and cannot be explained by the destruction of carbonate-containing planets to micron-sized grains. 
This leads \citet{Kemper_02_Detection} to conclude that an alternate formation mechanism for carbonates must exist that is not based on the weathering of rock in the presence of liquid water.  

In an accompanying paper, \citet{Kemper_02_mineral} present the data and describe the analysis that was done to come to this conclusion. 
The best fit is found through trial and error by varying parameters and visual comparison with the target (observed) spectrum, the so-called \emph{chi-by-eye} method. 
Here, we repeat this analysis, using \ampere\ to sample from the posterior distribution of the model parameters. 
\citet{Kemper_02_mineral} assume that the circumstellar environment is composed of a warm ($w$) and a cold ($c$) component, both in the form of spherically symmetric shells, and both contributing to the spectrum obtained with the Infrared Space Observatory (ISO):




\begin{equation}
\mathcal{F}_{\nu,\mathrm{w}, i} = \int_{T_\mathrm{out,w}}^{T_\mathrm{in,w}} \frac{1}{D^2} \frac{4 \pi a^2 {r_{0,\mathrm{w}}}^3 n_{0, i}}{3-p} \bigg( \frac{T(r)}{T_\mathrm{out,w}} \bigg)^{-\frac{3-p}{q}} Q_{\nu,\mathrm{w}, i} B_\nu(T) \, dT
\label{eq:fluxngc6302w}
\end{equation}

\begin{equation}
\mathcal{F}_{\nu,\mathrm{c}, i} = \int_{T_\mathrm{out,c}}^{T_\mathrm{in,c}} \frac{1}{D^2} \frac{4 \pi a^2 {r_{0,\mathrm{c}}}^3 n_{0, i}}{3-p} \bigg( \frac{T(r)}{T_\mathrm{out,c}} \bigg)^{-\frac{3-p}{q}} Q_{\nu,\mathrm{c}, i} B_\nu(T) \, dT
\label{eq:fluxngc6302c}
\end{equation}

\begin{equation}
\mathcal{F}_\nu = \sum_i^{n_{\rm w}}\mathcal{F}_{\nu,\mathrm{w}, i} + \sum_i^{n_{\rm c}}\mathcal{F}_{\nu,\mathrm{c}, i}
\label{eq:fluxngc6302}
\end{equation}

with $Q_{\nu,\mathrm{w}, i}$ and $Q_{\nu,\mathrm{c}, i}$ are the opacities of each species in the warm and cold components, respectively, where $i$ corresponds to the index of the species in the warm and cold component. 
The warm and cold components contain different subsets of the species, hence we separately sum over them in equation \ref{eq:fluxngc6302}.
Further, $D$ is the distance to the object, $a$ is the grain size, and $r_0$ is the inner radius of the temperature component. 
$p$ is the index of the power law of the density distribution with radius in each shell, and $q$ is the index of the power law of the temperature distribution with radius in each shell. 
The value $n_{0, i}$ is the density for each species at the inner radius of the shell, where the cold and warm shells are considered separately, and have different $n_{0, i}$ for the same species. 
Every dust component has its own value for $n_{0, i}$, and the ratios between these values yield the relative abundances. 

We have translated the original {\sc idl} code to calculate the specific flux into python. 
This code is included in {\tt ./examples/NGC6302.py}. 
We use the same opacity files as \citet{Kemper_02_Detection,Kemper_02_mineral}, which are stored in {\tt ./examples/NGC6302/}. 

We resolved the parameters in this model using \ampere\ {\tt ./examples/NGC6302.py} with the logarithm of the abundances ($n_0$), rather than the values themselves, as the values are very small but cannot be negative, and numerical errors are likely to occur otherwise. 
We used flat priors, with limits from -6 to 0 for the logarithm of the abundances, 10 and 80 K for the cold temperature component, and 80 and 180 K for the warm temperature component. 
The values for the indices $p$ and $q$ were fixed at $1/2$. The distance was set to 910 pc and the grain size to 0.1 \micron, following \citet{Kemper_02_mineral}. For the initial guess, the fit values from 2002 were used. 
The cold model component consists of amorphous olivine, forsterite, clino-enstatite, water ice, diopside, calcite and dolomite; while the warm component consists of amorphous olivine, metallic iron, forsterite, and clino-enstatite. 
\citet{Kemper_02_mineral} did not detect water ice, diopside, calcite and dolomite and placed upper limits on the abundance of these species in the warm component.

We ran \ampere\ with 50 walkers, and used 50000 samples. 
The burn-in value was set to 40000. Runs with smaller numbers of samples did not give satisfactory results because the abundances introduce strong correlations between parameters; this results in very long autocorrelation times for the chains. 
In order to achieve reasonable convergence, the parameter {\tt calUnc} was set to 10$^{-10}$, and {\tt scalelengthPrior} to 0.1 when defining the observed spectrum using the class {\tt Spectrum}. 
The fit was performed over the 12 to 120 \micron\ wavelength range, while the observed spectrum was resampled using linear spacing of 0.1 \micron\ up to 35 \micron; and a spacing of 0.0085 in $1/\lambda$ units at longer wavelengths. 
This sampling prevents fringing in the longer wavelength segments of the laboratory data from dominating the fit to the observed spectrum.

\begin{figure*}
    \centering
    \includegraphics[width=0.85\textwidth]{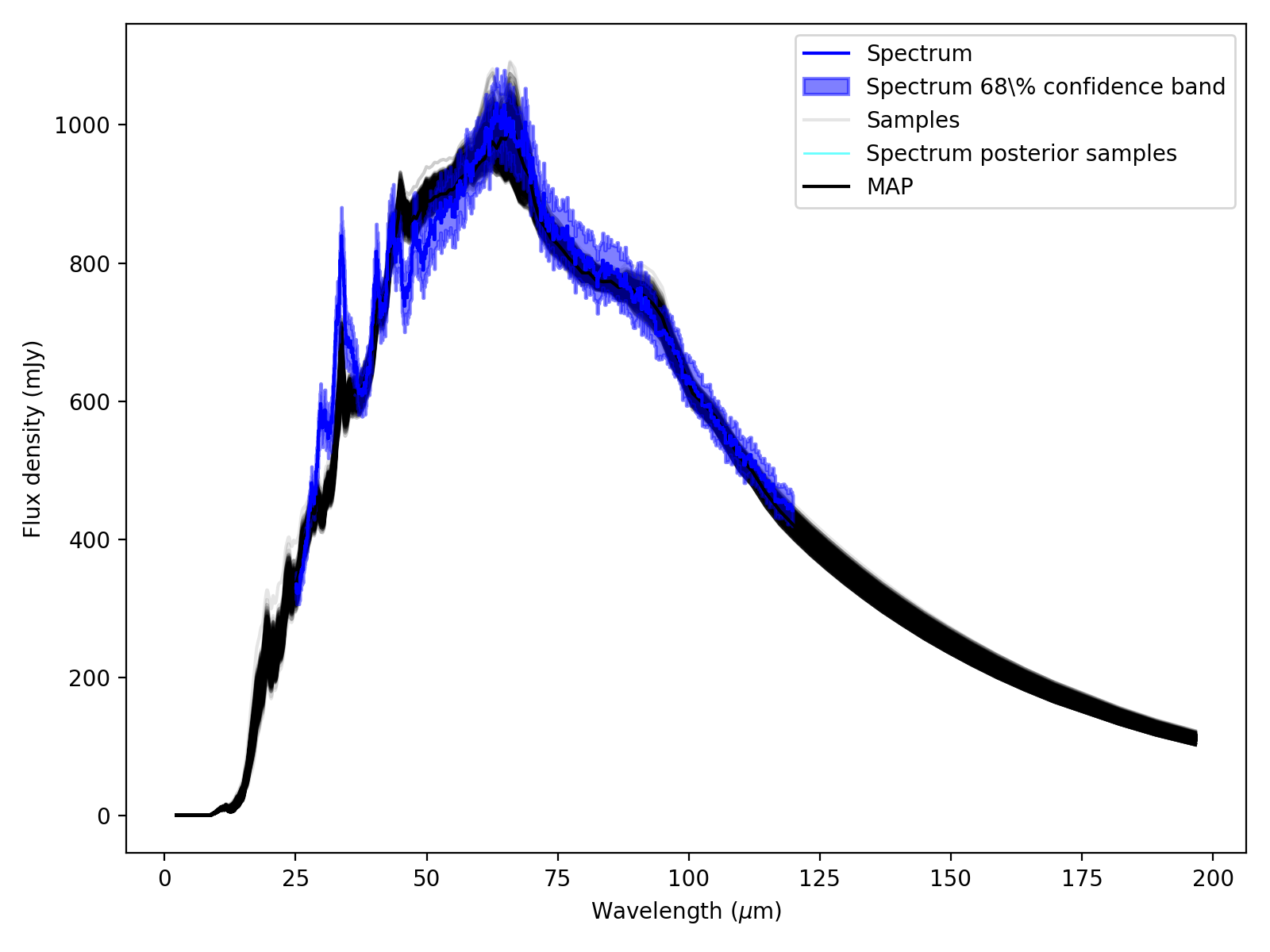}
    \caption{Output plot from \ampere\ for the NGC 6302 model. The blue line shows the observed spectrum plotted over the spectral range used to evaluate the fit. The blue shaded area shows the 68\% confidence band. The collection of black lines shows the various model outcomes.}
    \label{fig:NGC6302output}
\end{figure*}

The range of solutions shows a good resemblance to the observed spectrum (Fig.~\ref{fig:NGC6302output}; the corresponding output files are included in the file repository, and are labelled with a timestamp: {\tt 2022-12-19\_10-33-51}). 
The temperature ranges for the two temperature components are resolved as 57$\pm$2 to 36$\pm$2 K  and 123$\pm$3 to 105$\pm$3 K.   
In comparison, \citet{Kemper_02_Detection,Kemper_02_mineral} used 60 to 30 K and 118 to 100 K, and gave these numbers without error bars. 
Given that the priors allowed for a much broader range, the temperatures that were converged on with \ampere\ are remarkably close to, although not quite consistent with, the temperature ranges used by \citet{Kemper_02_mineral}.
The dust masses have been derived from $n_0$ using the script {\tt ./examples/NGC6302-calculate-dust-mass.py}.
When comparing the resulting dust composition (Tab.~\ref{tab:ngc6302}) with the composition derived by \citet{Kemper_02_mineral}, we see that the overall dust masses in the cold component are quite similar between both studies (0.048$^{+0.020}_{-0.010}$ M$_\odot$
for the current work, versus 0.050 M$_\odot$for \citet{Kemper_02_mineral}), the vast majority of which consists of amorphous olivine. We find we need a factor of four less cold forsterite in the present work, and other species in the cold component are even less abundant, compared to what \citet{Kemper_02_mineral} find. 
In particular, the mass contained in the species giving rise to the far-infrared part spectral signature -- calcite, dolomite, diopside and water ice -- is a factor of about ten lower in the present work (20 for water ice), but the features are still well reproduced in the solution. 
\citet{Kemper_02_mineral} cite an uncertainty of 50\% on the relative abundances, and a factor of 3 on the absolute dust masses, meaning that except for the overall dust mass and the relative abundances of amorphous and crystalline olivine, the results are not consistent with the present work.  
We thus conclude that while \ampere\ reproduces both the overall shape and the detailed characteristics of the spectrum very well, the determination of trace abundances deviates from the chi-by-eye approach. 
We consider the dust masses derived in the present work to be more robust, and they are also more comfortably reconciled with the available Calcium budget, compared to the work by \citet{Kemper_02_Detection, Kemper_02_mineral}.

\begin{table*} 
\caption{\label{tab:ngc6302} Dust masses in NGC 6302}
  \begin{tabular}{l c c c c c c}
\hline 
\hline
                     & \multicolumn{3}{c}{present work}                                                       & \multicolumn{3}{c}{\citet{Kemper_02_mineral}} \\
\hline
species              & $ M_{\mathrm{d,cold}}/M_{\odot}$      & mass fraction (cold) & $ M_{\mathrm{d,warm}}/M_{\odot}$  & $ M_{\mathrm{d,cold}}/M_{\odot}$   & mass fraction (cold) & $ M_{\mathrm{d,warm}}/M_{\odot}$\\
\hline
amorphous olivine    & $4.8^{+2.0}_{-1.0} \cdot 10^{-2}$   & 98.7\%                & $6.8^{+5.2}_{-3.1} \cdot 10^{-6}$ & $4.7 \cdot 10^{-2}$           & 94 \%                & $6.1 \cdot 10^{-6}$ \\
metallic iron        &                                  &                       & $1.5^{+1.1}_{-1.2} \cdot 10^{-5}$ &                              &                      & $1.2 \cdot 10^{-4}$ \\
forsterite           & $5.7^{+4.0}_{-1.6} \cdot 10^{-4}$   & 1.2\%                 & $8.7^{+7.3}_{-4.1} \cdot 10^{-8}$ & $> 2.0 \cdot 10^{-3}$         & $>$ 4.0 \%           & $3.7 \cdot 10^{-7}$ \\
clino-enstatite      & $3.9^{+5.0}_{-2.3} \cdot 10^{-6}$   & 0.008\%               & $8.8^{+6.4}_{-3.8} \cdot 10^{-8}$ & $> 5.5 \cdot 10^{-4}$         & $>$ 1.1 \%           & $3.1 \cdot 10^{-7}$ \\
water ice            & $1.5^{+1.7}_{-1.0} \cdot 10^{-5}$   & 0.03\%                &                              & $3.6 \cdot 10^{-4}$           & 0.72 \%              & $< 1.5 \cdot 10^{-8}$\\
diopside             & $3.0^{+1.7}_{-1.5} \cdot 10^{-5}$   & 0.06\%                &                              & $2.8 \cdot 10^{-4}$           & 0.56 \%              & $< 1.2 \cdot 10^{-7}$\\
calcite              & $1.5^{+0.8}_{-0.5} \cdot 10^{-5}$   & 0.03\%                &                              & $1.3 \cdot 10^{-4}$           & 0.26 \%              & $< 1.0 \cdot 10^{-7}$\\
dolomite             & $5.5^{+6.1}_{-3.2} \cdot 10^{-6}$   & 0.01\%                &                              & $7.9 \cdot 10^{-5}$           & 0.16 \%              & $< 3.0 \cdot 10^{-8}$\\
\hline
\hline
\end{tabular}
\end{table*}



\section{Summary}\label{sec:conc}
In this paper, we have demonstrated \ampere, a new package to interpret heterogeneous datasets.
We extend existing approaches to handle model misspecification based on flexible likelihood functions. 
\ampere\ implements general noise models which exploit kernel functions to produce a wide variety of correlated noise, making it straightforward to apply to a range of problems of interest to the astronomical community.
By combining these approaches with cutting-edge inference techniques, we demonstrate that even the most expensive astrophysical models, such as dust radiative transfer, can be used for Bayesian inference on large datasets.

We have demonstrated this with a series of examples representing common astronomical problems.
These problems include fitting a modified blackbody to photometry, determining stellar parameters from photometric and spectroscopic data, and calculating infrared excesses for circumstellar dust models. 

In each case, \ampere\ recovers values and uncertainties for the physical parameters of the model that are consistent with the `truth' provided by previous fitting using standard methods from the literature.
The comparison of \ampere\ with these previous analyses demonstrates its utility in obtaining robust estimates of parameters from model fitting.
In particular, \ampere\ illuminated the underlying or inherent biases in the previously adopted model fitting procedures that used limited data sets.
Further, \ampere\ revealed that uncertainties were generally underestimated by previous fitting procedures and in some cases substantially so (e.g. Sects \ref{sec:starplusdust} \& \ref{sec:cstar}). 
This last point is essential to the interpretation of large data sets particularly with the deployment of statistical techniques to infer the robustness of signal detection(s) amongst large samples.


\ampere\ is suitable for analyses on large samples, for example interpreting the large volume of spectra produced by missions such as JWST, SPHEREx, SDSS-V, and many more.
We are particularly interested in e.g., inferring the properties of evolved stars and their dust, to understand how they contribute to the chemical evolution of galaxies.
However, \ampere\ is much more widely applicable, and could for example be applied to problems as diverse as extracting the atmospheric parameters of exoplanets from transit spectroscopy, modelling high-redshift galaxy spectra to infer stellar populations or AGN fractions, or modelling emission-line spectra of ionised nebulae to infer abundances, temperatures and densities.

Future development will extend the existing framework with data classes appropriate for other widely used astrophysical data sets, such as (spectro-)imaging and interferometry. We also plan to integrate the popular JAX framework to accelerate \ampere, exploiting just-in-time compilation, GPU acceleration, and automatic differentiation. The modular nature of the Model classes and interaction with a wide variety of Inference tools through a framework of common interfaces enables \ampere\ to be flexible and extensible to the needs of users and future software developments. We look forward to working with the wider astronomical community to realise the potential of \ampere, and hope that others will join us as users, collaborators and fellow developers.


\section*{Acknowledgements}
All the authors previously worked together at Academia Sinica Institute of Astronomy and Astrophysics (ASIAA). We would like to thank ASIAA for bringing us together and supporting the early stages of \ampere's development.

JPM, FK and LP acknowledge support by the Ministry of Science and Technology of Taiwan under grant MOST 107-2119-M-001-031-MY3, and Academia Sinica under grant AS-IA-106-M03. 
JPM acknowledges research support by the National Science and Technology Council of Taiwan under grants NSTC 109-2112-M-001-036-MY3 and NSTC 112-2112-M-001-032-MY3.
SS acknowledges support from UNAM-PAPIIT Programs IA 104822 and IA 104824.
This work was also partly supported by the Spanish program Unidad de Excelencia Mar{\'{\i}}a de Maeztu CEX2020-001058-M, financed by MCIN/AEI/10.13039/501100011033.
FI-B-00617. S.Z. acknowledges support from the European Space Agency (ESA) as an ESA Research Fellow.
L.F. acknowledges support from the Ministry of Science and Technology / National Science and Technology Council of Taiwan under grants no. 111-2811-M-005-007, 109-2112-M-005-003-MY3, and 114-2811-M-032-005.

This work has made use of the Python programming language \citep{python_CS-R9526}, and the packages {\sc Astropy} \citep{2013AstroPy,2018AstroPy}, {\sc Corner} \citep{2016ForemanMackey}, {\sc dynesty} \citep{dynesty_v1_paper}, {\sc Emcee} \citep{emcee_paper}, {\sc Extinction} \citep{Extinctionpkg}, {\sc Hyperion} \citep{Hyperion_paper} , {\sc Matplotlib} \citep{2007Hunter}, {\sc MiePython}, {\sc NumPy} \citep{2020Harris}, {\sc pyphot} \citep{pyphot}, {\sc sbi} \citep{tejero-cantero2020sbi}, {\sc SciPy} \citep{2020Virtanen}, {\sc Spectres} \citep{Spectres_paper}, {\sc Starfish} \citep{Starfish_paper}, and {\sc Zeus} \citep{karamanis2021zeus}.

\section*{Data Availability}
All data in this work is either in the public domain, or can be generated with scripts available at \url{https://github.com/ICSM/ampere/tree/master/examples}.

\bibliographystyle{aasjournal}
\bibliography{biblio.bib} 




\appendix

\section{Corner plots}

\begin{figure*}
    \centering
    \includegraphics[width=0.75\textwidth]{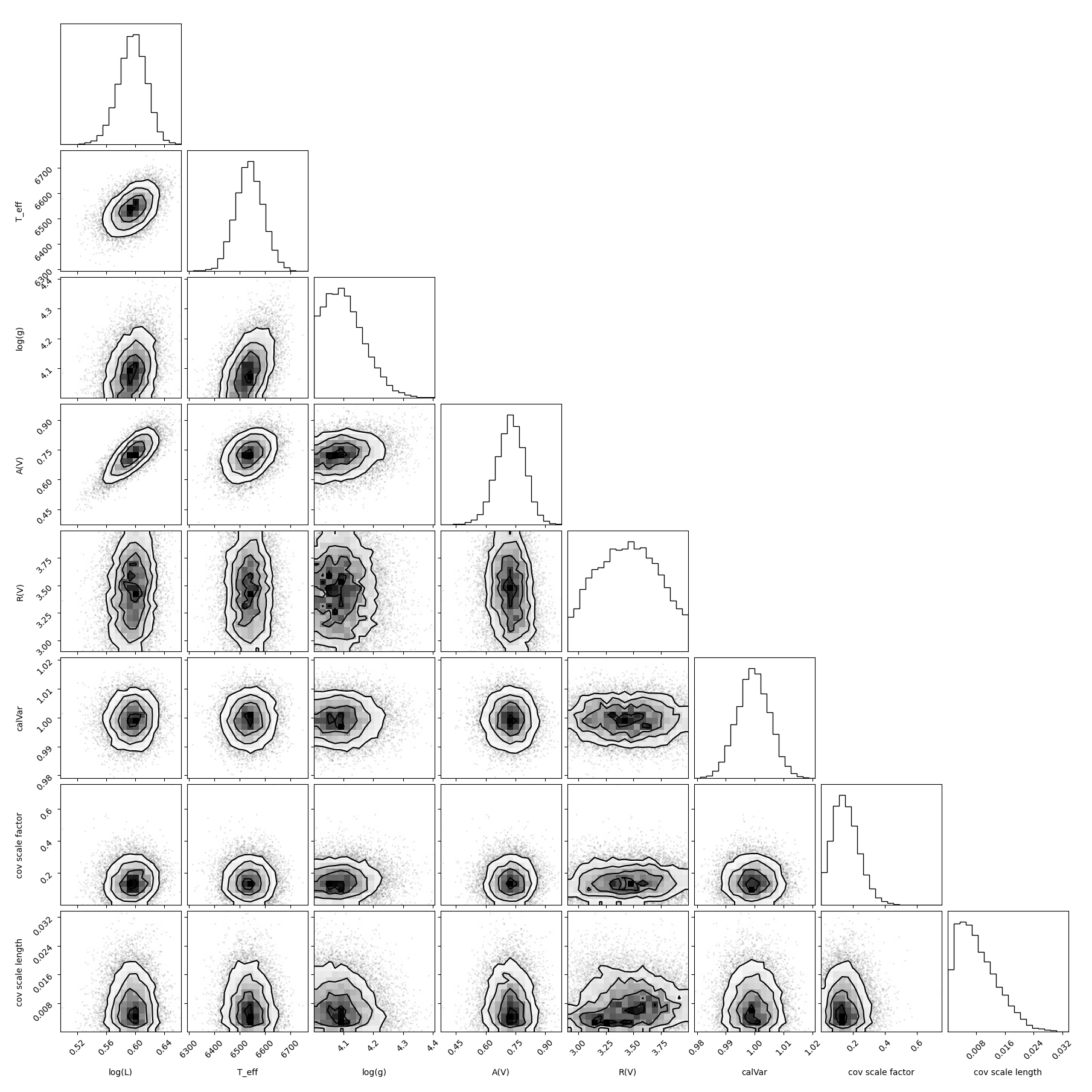}
    \caption{Corner plot for photometry+spectrum inference of astrophysical parameters. As expected, there are mild correlations between $\log\left(L\right)$, $T_{\rm eff}$ and $A_{\rm V}$.}
    \label{fig:corneraps}
\end{figure*}

\begin{figure*}
    \centering
    \includegraphics[width=0.9\textwidth]{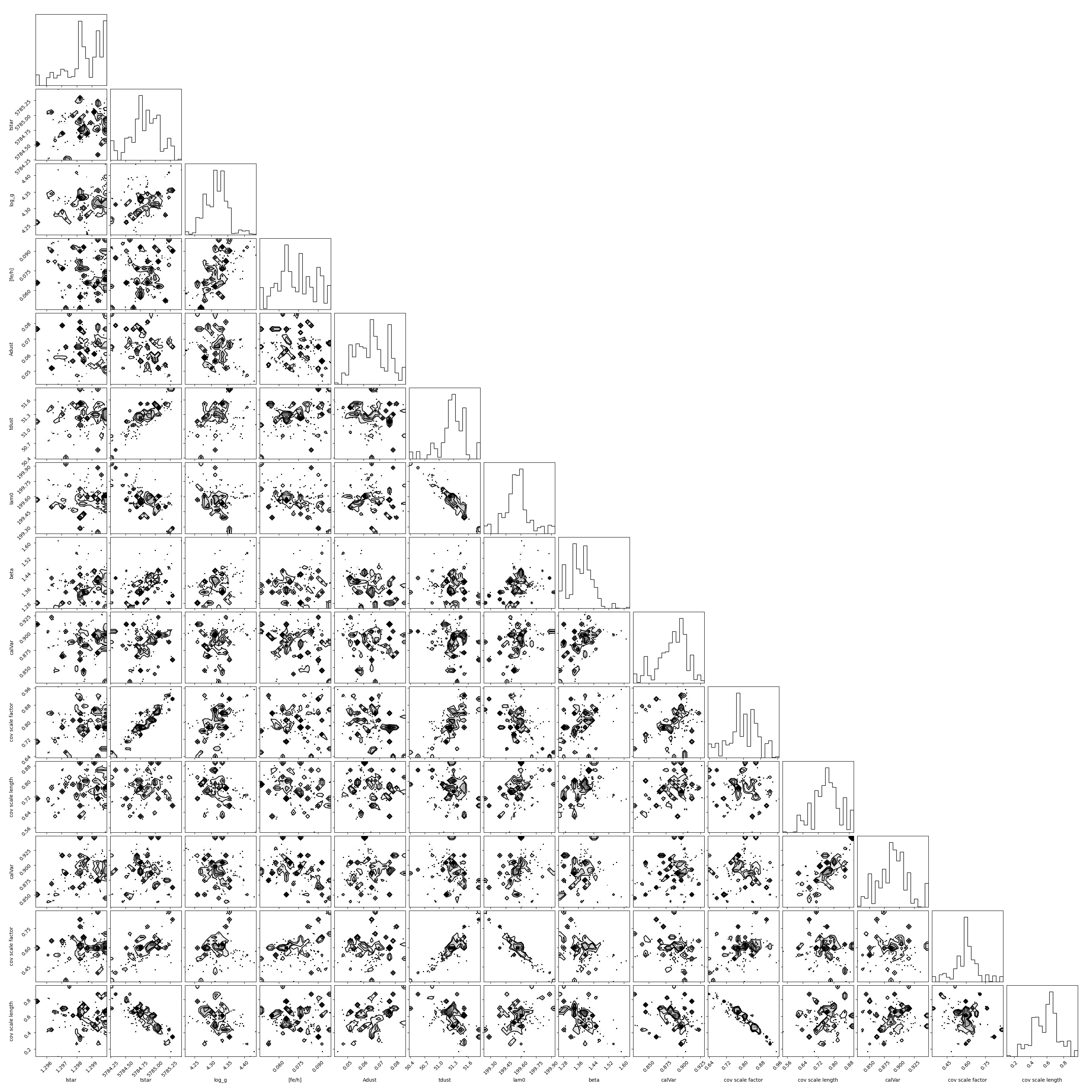}
    \caption{Corner plot for the star plus disc model showing posterior probability distribution of the model fitting as described in the text. The posterior probability distributions are comprised of 6,000 realisations of the model.}
    \label{fig:starndust_corner}
\end{figure*}

\begin{figure*}
    \centering
    \includegraphics[width=0.75\textwidth]{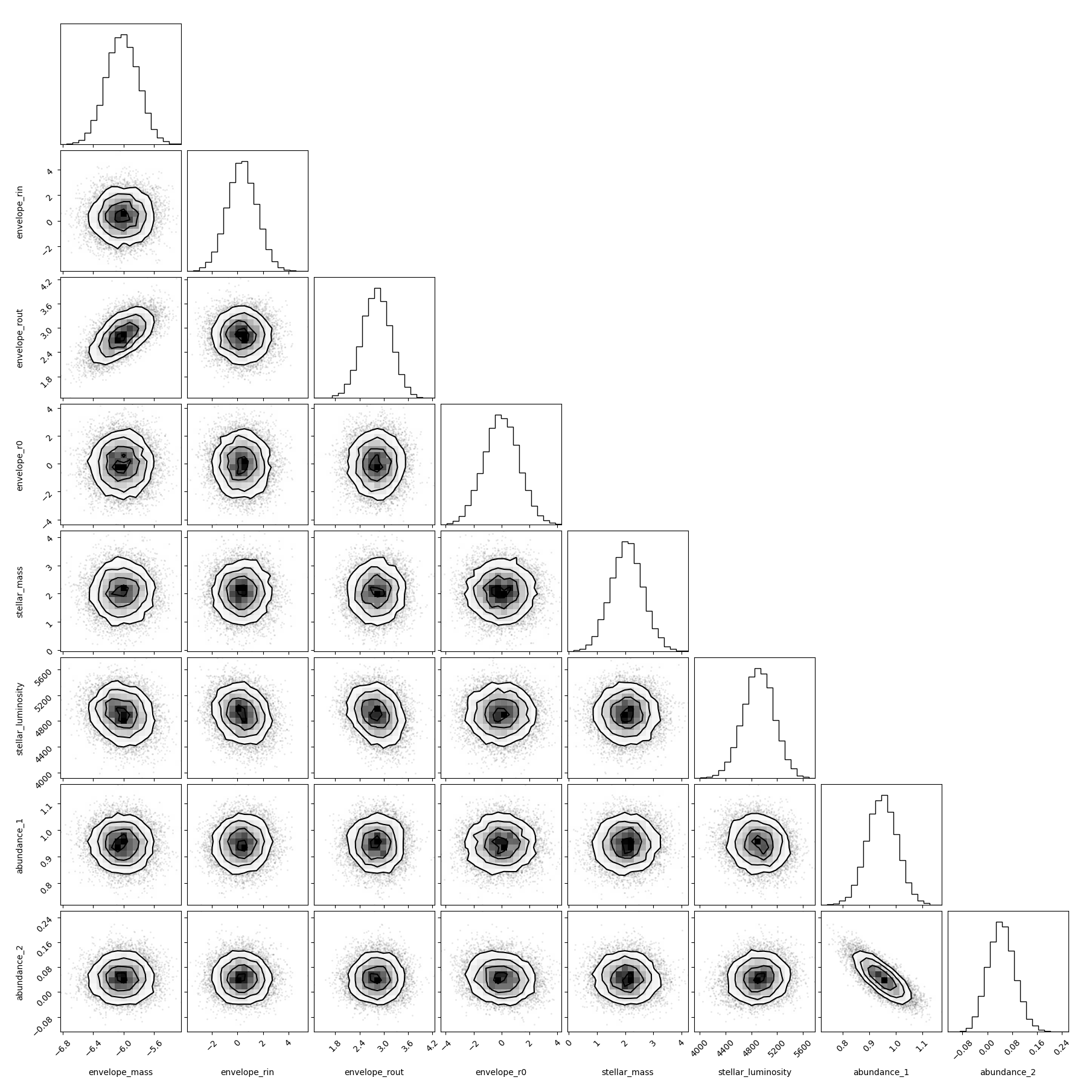}
    \caption{Corner plot for the carbon-star model, showing only the parameters of the physical model and excluding the noise-model parameters.}
    \label{fig:cstar-corner}
\end{figure*}



\label{lastpage}
\end{document}